\documentclass[11pt]{article}
\usepackage{amsmath}
\usepackage{amsfonts}

\def\be{\begin{equation}}
\def\ee{\end{equation}}
\def\bea{\begin{eqnarray}}
\def\eea{\end{eqnarray}}

\def\scri{\mathcal{I}}

\def\hO{\hat{O}}
\def\hE{\hat{E}}
\def\hF{\hat{F}}

\begin{document}

\title{Valiente Kroon's obstructions to smoothness at infinity}
\author{James Grant\thanks{email: j.grant@surrey.ac.uk}\\Department of Mathematics,\\
 University of Surrey,	
\\ \\Paul Tod\thanks{email: tod@maths.ox.ac.uk}\\Mathematical Institute\\University of Oxford}

\maketitle
\begin{abstract}
We conjecture an interpretation in terms of multipole moments of the obstructions to smoothness at infinity found for time-symmetric, conformally-flat initial data by Valiente Kroon \cite{vk1}.
\end{abstract}
\section{Introduction}
In \cite{vk1} Valiente Kroon identified a new class of obstructions to smoothness at infinity for asymptotically-flat solutions of the Einstein vacuum equations. He used the formalism of Friedrich \cite{f1} which begins
by identifying universal structure at null and spatial infinity and then writes out a system of conformal Einstein equations, equivalent to the vacuum equations, with adapted spin frames and coordinates. Friedrich \cite{f1}
found obstructions to smoothness at infinity in these coordinates in the time-symmetric case: necessary conditions for smoothness at infinity in the time-symmetric case are the vanishing at infinity of the symmetrised spinor derivatives of all orders of the spinor corresponding to the Bach tensor;
then finite degrees of smoothness
require finite numbers of these derivatives to vanish.

Valiente Kroon \cite{vk1} considered time-symmetric initial data with the extra condition of conformal-flatness, that is the second fundamental form is zero and the 3-metric of the data surface is conformally-flat, which entails the vanishing of the
Bach tensor. Thus Friedrich's conditions on the Bach tensor are vacuously satisfied but, by careful analysis of Friedrich's system, Valiente Kroon is able to show that there are still obstructions to smoothness. He finds a
heirarchy of obstructions
labelled by quantities $G^{(n)}$ for $n\geq 5$ and where each $G^{(n)}$ is a set of $2n-5$ constants obtained from a spherical harmonic expansion of a harmonic function obtained from the conformal factor to flat-space. He observes
that the first obstruction $G^{(5)}$ is essentially the Newman--Penrose conserved quantity for the gravitational field \cite{np1} (see also \cite{tb1}, \cite{dvk}). In this note, we give reasons for believing that in general
Valient Kroon's obstructions are proportional to a family of tensors which generalise the NP quantities as they relate to multipoles. That there is a connection between his obstructions and multipoles is implicit in the work of
Valiente Kroon, and explicit in the proof of rigidity \cite{vk2}, but the explicit relation conjectured here is thought to be new.

Time-symmetric, conformally-flat initial data then are determined by a 3-metric
\be\label{mm1}h_{ij}dx^idx^j=V^4\delta_{ij}dx^idx^j,\ee
where the constraint equations for the Einstein vacuum equations reduce to the Laplace equation on $V$. Asymptotic flatness can be achieved by $V\rightarrow 1$ at large distances.
Commonly considered solutions (e.g. \cite{bl}) are sums of point masses
\be\label{h1}
V=1+\frac{1}{2}\sum_\alpha\frac{m_\alpha}{|{\bf{r}}-{\bf{a}}_\alpha|}.
\ee
Here the factor $1/2$ before the sum is to ensure that the total (ADM) mass at infinity is $M=\sum m_\alpha$.

As is familiar, the singularities of $V$ at ${\bf{r}}={\bf{a}}_\alpha$ are not singularities of $h_{ij}$ but rather are other `ends' in the sense of asymptotically-flat regions, \cite{bl}. Also one may construct solutions like
the Misner wormhole \cite{mis} by superposing a suitable infinite set of collinear point masses.

The outline of this letter is as follows: in the next section we review the definition of multipole moments in ${\mathbb{R}}^3$ and introduce a set of trace-free, symmetric tensors $\hE^{(n)}_{i_1\ldots i_{n+1}}$ derived from
the multipole moments. In Section 3 we recall Valiente Kroon's obstructions to smoothness at infinity \cite{vk1} and conjecture a relation between them and the tensors $\hE^{(n)}$. In Section 4 we say a few words about the
issue of rigidity in the sense that vanishing of the obstructions constrains the data to be data for Schwarzschild.

\section{Multipole moments in flat space}
Given a harmonic function $V$ in ${\mathbb{R}}^3$ tending to one at large distances, as in the previous section, one has an expansion in spherical harmonics
\[V = 1+\sum_{\ell=1}^\infty\sum_{m=-\ell}^{\ell}r^{-(\ell+1)}c_{\ell m}Y_{\ell m}(\theta,\phi)\]
in the standard spherical polar coordinates. This can alternatively be written as
\[V=1+\sum_{\ell=1}^\infty r^{-(\ell+1)}m_{i_1\ldots i_\ell}e_{i_1}\ldots e_{i_\ell},\]
where the summation convention applies to tensor indices, we introduce the vector ${\bf{e}}=(e_i)=(\sin\theta\cos\phi,\sin\theta\sin\phi,\cos\theta)$ and for each $\ell$ the tensor $m_{i_1\ldots i_\ell}$ is trace-free and symmetric.
For each $\ell$, there are $2\ell+1$ coefficients $c_{\ell m}$ and the tensor $m_{i_1\ldots i_\ell}$ has $2\ell+1$ independent components, so that each can be written uniquely in terms of the other. Following varying conventions these components either are
or are proportional to the $(\ell+1)$-th or $2^\ell$-th multipole moment.

If $V$ solves a Poisson equation
\[\nabla^2V=\kappa\rho\]
with density $\rho$ then the multipole moments can be related to integrals over the source in a familiar manner. The case of interest for us is when the source consists of a set of
point masses of mass $m^\alpha$ at positions ${\bf{a}}^\alpha=(a^{\alpha}_i)$ in ${\mathbb{R}}^3$ with $\alpha$ ranging over some indexing set. It is convenient to define
\be\label{m1}
O^{(n)}_{i_1\ldots i_n}=\sum_{\alpha}m^\alpha a^\alpha_{i_1}\ldots a^\alpha_{i_n},
\ee
with $O^{(0)}=M$, the total mass. Expansion of $V$ in (\ref{h1}) now shows that the tensor expression of the $n$-th multipole moment is proportional to the tensor
\be\label{m2}
\hO^{(n)}_{i_1\ldots i_n}=O^{(n)}_{i_1\ldots i_n}-\mbox{  trace}
\ee
where the subtracted trace is an expression that we don't need explicitly, made from a combination of Kronecker deltas and all possible traces of $O^{(n)}$. It is computationally simpler to work with
$O^{(n)}$ but keep in mind that it is the symmetric and trace-free $\hO^{(n)}$ which corresponds to the multipole.

We obtain a compression of notation by use of polynomials: introduce $(X^i)=(X,Y,Z)$ and then symmetric tensors $t_{i_1\ldots i_k}$ are in one-to-one correspondence with polynomials of degree $k$ via
\[t(X,Y,Z):=t_{i_1\ldots i_k}X^{i_1}\ldots X^{i_k}.\]
In particular this gives $n$-th order polynomials $O^{(n)}$ and $\hO^{(n)}$ representing the tensors in (\ref{m1}) and (\ref{m2}).

Note that the tensor $\delta_{ij}$ corresponds to the polynomial $r^2:=X^2+Y^2+Z^2$. Call a tensor $t_{i_1\ldots i_n}$ \emph{pure trace} if it is of the form $\delta_{(i_1i_2}s_{i_3\ldots i_n)}$ for some $s_{i_3\ldots i_n)}$,
since it will then have a zero trace-free part. As polynomials this condition is
\be\label{tf}t^{(n)}=r^2s^{(n-2)},\ee
which we use below.

Under shift of origin by $b_i$ in ${\mathbb{R}}^3$ the point masses move according to:
\[a^{\alpha}_{i}\rightarrow a^\alpha_i+b_i,\]
which can be written in terms of polynomials as:
\[a^\alpha\rightarrow a^\alpha+b.\]
Now introduce $\delta$ for the change under translation so that
\[\delta a^\alpha=b\]
and deduce from (\ref{m1}) that
\be\label{s1}
\delta O^{(n)}=\sum_{k=1}^n\,^{n}C_kb^kO^{(n-k)}.
\ee
For $n\geq 1$, introduce an infinite set of tensors
\be\label{e1}
E^{(n)}_{i_1\ldots i_{n+1}}:=M^{n}O^{(n+1)}_{i_1\ldots i_{n+1}}-O^{(1)}_{(i_1}\ldots O^{(1)}_{i_{n+1})}
\ee
so that, as polynomials
\[E^{(n)}=M^{n}O^{(n+1)}-(O^{(1)})^{n+1}\]
and then write $\hE^{(n)}$ 
for the trace-free part of $E^{(n)}$.

It is straightforward to see that, under translation of origin,  $E^{(n)}$ transforms according to
\be\label{e2}
\delta E^{(n)}=\sum_{k=1}^{n-1}\,^{n+1}C_kM^kb^kE^{(n-k)}
\ee
for $n>1$ while $E^{(1)}$ is independent of origin, and the higher $E^{(n)}$ transform in terms of the lower ones.

Since taking the trace commutes with $\delta$, $\hE^{(1)}$ is also origin-independent and it can in fact be identified as the tensor
defining the NP conserved quantities (see e.g. \cite{dvk}, \cite{vk1}).

From (\ref{e2}) we also deduce that
\begin{itemize}
 \item if $E^{(k)}=0$ for $1\leq k<n$  then $E^{(n)}$ is origin-independent.
\end{itemize}
This result also holds for $\hE^{(n)}$ by the following argument: suppose $\hE^{(k)}=0$ for $1\leq k<n$ then for this range $E^{(k)}$ is pure trace i.e. by (\ref{tf})
\[E^{(k)}=r^2S^{(k-2)}\]
for some $S^{(k-2)}$. From (\ref{e2}) it follows that $\delta E^{(n)}$ is also pure trace and therefore, since taking the trace commutes with $\delta$, that $\delta\hE^{(n)}=0$.


We may choose the origin to set $O^{(1)}=0$, and this choice defines the centre-of-mass frame. In the centre-of-mass frame by (\ref{e1}) each $E^{(n)}$ is proportional to the corresponding $O^{(n)}$, and therefore each
$\hE^{(n)}$ is proportional to the corresponding $\hO^{(n)}$. Thus
\begin{itemize}
 \item in the centre-of-mass frame if $\hE^{(k)}=0$ for $1\leq k\leq n$ then $\hO^{(k)}=0$ for $1\leq k\leq n$, and vice versa.
\end{itemize}
whence also
\begin{itemize}
 \item if $\hE^{(k)}=0$ for all $k$ then in the centre-of-mass frame $\hO^{(k)}=0$ for all $k>0$ and $V$ is spherically symmetric.
\end{itemize}
It is worth noting that there is another set of tensors with properties similar to the $E^{(n)}$ and $\hE^{(n)}$, namely $F^{(n)}$ and $\hF^{(n)}$ where, defined as polynomials:
\be\label{fn}
F^{(n)}:=MO^{(n+1)}-O^{(1)}O^{(n)},\ee
for $n\geq 1$, and $\hF^{(n)}$ is the trace-free part of $F^{(n)}$. In place of (\ref{e2}) one finds
\be\label{f2}
\delta F^{(n)}=\sum_{k=1}^{n-1}\,^{n}C_kb^kF^{(n-k)},
\ee
and the discussion proceeds as before. Evidently one could express the $E^{(n)}$ in terms of the $F^{(n)}$ and vice versa, but it is the $E^{(n)}$ which arise most immediately in Valiente Kroon's obstructions.

It remains to be seen whether the $\hE^{(n)}$ or $\hF^{(n)}$ have significant properties at null infinity, $\scri^+$.

\section{Valiente Kroon's obstructions}
In \cite{vk1} Valiente Kroon studied Friedrich's system \cite{f1} for time-symmetric, conformally-flat initial data. The data are the metric in the form (\ref{mm1}). He was able to integrate the equations at infinity for the first
four orders, giving smooth solutions, but logarithmic terms arise at the intersection of the cylinder at space-like infinity with null infinity at all orders from the fifth order and will obstruct smoothness at infinity. The coefficients of the logarithms at order $n$ for $n\geq 5$ are
quantities $G^{(n)}$, where each $G^{(n)}$ is a set of $2n-5$ constants expressed in terms of the coefficients of the expansion of $V$ in spherical harmonics.

Valiente Kroon expanded $V$ in terms of functions $T^{\;\;k}_{n\;\;\;j}$ which are a complete orthonormal set for $L^2(SU(2,{\mathbb{C}}))$ with standard Haar measure, and are standardly used by Friedrich and his collaborators (see e.g.
\cite{f1}, \cite{fk}). With a suitable choice of conventions, they can be related to spin-weighted spherical harmonics up to numerical constants by
\[T^{\;\;k}_{n\;\;\;j}\sim e^{is\psi}\,_sY_{\ell m}(\theta,\phi),\]
with $s=\frac12n-j,\ell=\frac12n,m=\frac12n-k$ and Euler angles $(\theta,\phi,\psi)$. In particular
\[T^{\;\;k}_{2\ell\;\;\;\ell}\sim Y_{\ell m}\]
with $m=\ell-k$. If one is more familiar with spherical harmonics, this observation simplifies the interpretation of Valiente Kroon's obstructions. His $G^{(5)}$ contains coefficients $w_{2,4,k}$ and $w_{1,2,k}$: $w_{2,4,k}$ is the
coefficient of $T^{\;\;k}_{4\;\;\;2}$ in $V$, and therefore is related to the coefficient of $Y_{2,2-k}$ and therefore is a component of the tensor $\hO^{(2)}$; $w_{1,2,k}$ is similarly related to $\hO^{(1)}$;
so $G^{(5)}$ is a sum of two terms proportional respectively to  $M\hO^{(2)}$ and the trace-free part of $(O^{(1)})^2$. It is a trace-free tensor under rotation and, as the leading obstruction, must be origin-independent. Therefore
it must be $\hE^{(1)}$, a fact shown explicitly in \cite{vk1}, but this model of argument can be applied at higher orders.

Moving on to $G^{(6)}$, each component is a sum of two kinds of term, one obtained from $M^2\hO^{(3)}$ and the other from the trace-free part of $(O^{(1)})^3$. Again it's a trace-free tensor under rotation and
if $G^{(5)}=0$ it becomes the leading obstruction and therefore origin-independent. Thus it must be proportional to $\hE^{(2)}$. Inductively we are lead to conjecture that, for all $k\geq 5$, $G^{(k)}$ and $\hE^{(k-4)}$ are proportional.

\section{Rigidity}
Valiente Kroon \cite{vk2} shows that smoothness at infinity implies that the data is that for the Schwarzschild solution. This would follow from the conjecture in the preceding
section since the vanishing of all $G^{(k)}$ entails the vanishing of all $\hE^{(k)}$ and hence, in the centre-of-mass frame, of all multipole moments $\hO^{(n)}$ for $n\geq 1$ which implies Schwarzschild.
One can obtain various intermediate results:

\begin{enumerate}
\item[$\bullet$] in the case that all ${\bf{a}}_\alpha$ are coplanar then the vanishing of $E^{(1)}$ implies Schwarzschild (i.e. all the mass points are coincident); in particular this also holds for coplanar Misner wormholes;
\item[$\bullet$] with 4 mass-points there is a one-parameter family of tetrahedral configurations with $E^{(1)}=0$ which are equilateral if the masses are all equal; however
the vanishing of $E^{(2)}$ forces Schwarzschild;
\item[$\bullet$] so one is lead to conjecture: for each $n$ there will be an $N_n$ such that the vanishing of $\hE^{(k)}$ for $1\leq k\leq n$ forces any configuration of $j\leq N_n$ mass points to be Schwarzschild. A naive count
of equations suggests $3N_n$ is the largest multiple of 3 less than $n^2+4n+7$.
\end{enumerate}

\section*{Acknowledgement}
We are grateful to Prof Valiente Kroon for discussions. JG is grateful to St. John's College, Oxford for a Visiting Scholarship during which this work was carried out. He is also grateful to the Mathematical Institute in Oxford for their hospitality.


\end{document}